\documentclass[twocolumn,aps,prl,a4paper,amsmath,amsfonts,10pt]{revtex4-2}

\pdfoutput=1

\usepackage[compat=0.4]{yquant}
\usepackage{braket,xcolor,hyphenat,microtype,adjustbox,ragged2e}
\usepackage[colorlinks, linkcolor=linkblue]{hyperref}
\usepackage[english]{babel}
\usepackage[frozencache]{minted}
\usetikzlibrary{quotes,fit}
\useyquantlanguage{qasm,groups}

\makeatletter

\begingroup
\catcode`\&=11

\endgroup

\definecolor{boxGrayBody}{gray}{.94}
\colorlet{linkblue}{cyan!35!blue}
\setminted{frame=none, linenos=false, fontsize=\footnotesize, breaklines, bgcolor=boxGrayBody, samepage}
\setmintedinline{bgcolor={}, fontsize=}

\newmintinline[tex]{tex}{}
\newsavebox\codeexamplebox
\newenvironment{codeexample}{%
   \VerbatimEnvironment%
   \let\FVB@VerbatimOut\minted@FVB@VerbatimOut
   \let\FVE@VerbatimOut\minted@FVE@VerbatimOut
   \minted@configlang{tex}%
   \minted@fvset
   \begin{VerbatimOut}[codes={\catcode`\^^I=12},firstline,lastline]{\minted@jobname.pyg}%
}{
   \end{VerbatimOut}%
   \minted@langlinenoson%
   \savebox\codeexamplebox{\input \minted@jobname.pyg}%
   \ifdim\wd\codeexamplebox>\dimexpr.5\linewidth-3mm\relax%
      \wd\codeexamplebox=.5\linewidth
   \else%
      \wd\codeexamplebox=\dimexpr\wd\codeexamplebox+3mm\relax
   \fi%
   \noindent\begin{minipage}{\wd\codeexamplebox}%
      \centering%
      \usebox\codeexamplebox%
   \end{minipage}%
   \begin{minipage}{\dimexpr\linewidth-\wd\codeexamplebox\relax}%
      \minted@pygmentize{\minted@lang}%
   \end{minipage}%
   \minted@langlinenosoff%
   \par%
}

\newenvironment{codeexample*}{%
   \VerbatimEnvironment%
   \let\FVB@VerbatimOut\minted@FVB@VerbatimOut
   \let\FVE@VerbatimOut\minted@FVE@VerbatimOut
   \minted@configlang{tex}%
   \minted@fvset
   \begin{VerbatimOut}[codes={\catcode`\^^I=12},firstline,lastline]{\minted@jobname.pyg}%
}{
   \end{VerbatimOut}%
   \minted@langlinenoson%
   \begin{adjustbox}{center}
      \input \minted@jobname.pyg %
   \end{adjustbox}\nopagebreak
   \minted@pygmentize{\minted@lang}%
   \minted@langlinenosoff%
   \par%
}

\newenvironment{codeexample**}{%
      \VerbatimEnvironment%
      \let\FVB@VerbatimOut\minted@FVB@VerbatimOut
      \let\FVE@VerbatimOut\minted@FVE@VerbatimOut
      \minted@configlang{tex}%
      \minted@fvset
      \begin{VerbatimOut}[codes={\catcode`\^^I=12},firstline,lastline]{\minted@jobname.pyg}%
}{
      \end{VerbatimOut}%
      \minted@langlinenoson%
      \begin{adjustbox}{center}
         \input \minted@jobname.pyg %
      \end{adjustbox}
   \endwidetext%
   \minted@pygmentize{\minted@lang}%
   \minted@langlinenosoff%
   \par%
}

\newcommand{\qpic}{$\langle\mathsf{q}|\mathsf{pic}\rangle$}
\newcommand{\TikZ}{Ti\textit kZ}

\begin{document}
   \frenchspacing
   \title{yquant: Typesetting quantum circuits in a human-readable language}
   \author{Benjamin Desef}
   \affiliation{Institute for Theoretical Physics, University of Ulm}
   \date{\today}

   \begin{abstract}
      yquant is a \LaTeX-only package---i.e., involving no external programs---that allows to typeset quantum circuits in a simple and human-readable language.
      It is inspired by the well-known qasm language and in fact even supports this language; however, the yquant language brings much more flexibility.
      yquant is based on \TikZ{} and, while producing high-quality output out\hyp of\hyp the\hyp box, it is highly configurable.
   \end{abstract}

   \maketitle

   \section{Introduction}
      There are already various \TeX{} packages available that allow to typeset quantum circuits.
      They can in general be divided into two categories: Packages that use external (usually Python) scripts will translate a rather well\hyp understandable text file that specifies the quantum circuit into \TeX{} commands.
      Among them is qasm~\cite{bib:qasm}, probably the first of these packages, which was also used to produce the circuits in Nielsen and Chuang's famous book on quantum information~\cite{bib:Nielsen2010}; and also \qpic~\cite{bib:qpic}.

      The other kind of packages makes the creation of quantum circuits in pure \LaTeX{} easier by providing a set of macros that produce the required shapes, but the main job of casting the ``logical'' circuit into a tabular layout has to be done by the author.
      Among them is qcircuit~\cite{bib:qcircuit} and quantikz~\cite{bib:quantikz}.

      yquant combines both approaches into one: It is a pure \LaTeX{} solution that nevertheless allows the circuits to be entered in a human-readable language.
      It internally produces \TikZ~\cite{bib:tikz} commands and in fact, arbitrary interaction with the \TikZ{} layer is possible for manual intervention.
      yquant will automatically determine all necessary positioning, something that external script usually cannot do well---since they do not know how large the final elements will be.

      The yquant language is made to be particularly easy to read, being thus more verbose than \qpic.
      It at the same time brings an enormous flexibility and great versatility with regard to register addressing.

      Not only does yquant bring its own language; it also comes with a module that allows to directly read qasm syntax, which eases the transition from qasm to yquant.

      yquant is licensed under the \LaTeX{} Project Public License~1.3c; its source code is freely available on GitHub~\cite{bib:yquantsrc}, and it is on the CTAN.

      This announcement of yquant on the arXiv is only a brief overview and summary; for more details and more examples, refer to the yquant package documentation~\cite{bib:yquantdoc}.

   \widetext
   \section{Examples}
      We load yquant by \tex!\usepackage[compat=0.4]{yquant}!; these examples additionally use the \texttt{braket} package.
      \TikZ{} is automatically loaded by yquant.

      \subsection*{Simple teleportation circuit}
         \begin{codeexample}
\begin{tikzpicture}
   \begin{yquant}
      qubit {$\ket{\reg_{\idx}}$} q[3];

      h q[1];
      cnot q[2] | q[1];
      cnot q[1] | q[0];
      h q[0];
      measure q[0-1];
      z q[2] | q[1];
      x q[2] | q[0];
   \end{yquant}
\end{tikzpicture}
         \end{codeexample}

      \subsection*{Phase estimation}\vspace*{-4mm}
         \begin{codeexample}
\begin{tikzpicture}
   \begin{yquant}
      qubit {$\ket{j_{\idx}} = \ket0$} j[3];
      qubit {$\ket{s_{\idx}}$} s[2];

      h j;
      box {$U^4$} (s) | j[0];
      box {$U^2$} (s) | j[1];
      box {$U$} (s) | j[2];
      h j[0];
      box {$S$} j[1] | j[0];
      h j[1];
      box {$T$} j[2] | j[0];
      box {$S$} j[2] | j[1];
      h j[2];
      measure j;
   \end{yquant}
\end{tikzpicture}
         \end{codeexample}

      \subsection*{FT error correction}\vspace*{-4mm}
         \begin{codeexample*}
\begin{tikzpicture}
   \begin{yquant}
      qubit {$\ket{q_{\idx}}$} q[3];
      qubit {$\ket{s_{\idx}} = \ket0$} s[2];
      cbit {$c_{\idx} = 0$} c[2];

      h s[0];
      cnot s[1] | s[0];
      cnot s[0] | q[0];
      cnot s[1] | q[1];
      cnot s[1] | s[0];
      h s[0];
      measure s;
      cnot c[0] | s[0];
      cnot c[1] | s[1];
      discard s;

      init {$\ket0$} s;
      h s[0];
      cnot s[1] | s[0];
      cnot s[0] | q[1];
      cnot s[1] | q[2];
      cnot s[1] | s[0];
      h s[0];
      measure s;

      box {Process\\Syndrome} (s, c);
      box {$\mathcal R$} (q) | s, c;
   \end{yquant}
\end{tikzpicture}
         \end{codeexample*}

      \subsection*{Error correction/Use of subcircuits}
         \emph{Subcircuits are a new feature of yquant~0.2.}

         \begin{codeexample}
% \usetikzlibrary{quotes}
\begin{tikzpicture}
  \begin{yquant}
    qubit {} msg[3];
    nobit syndrome[3];

    [this subcircuit box style={dashed, "Syndrome Measurement"}]
    subcircuit {
       qubit {} msg[3];
       [out]
       qubit {$\ket0$} syndrome[3];

       cnot syndrome[0] | msg[0];
       cnot syndrome[0] | msg[1];
       cnot syndrome[1] | msg[1];
       cnot syndrome[1] | msg[2];
       cnot syndrome[2] | msg[0];
       cnot syndrome[2] | msg[2];

       dmeter {$M_{\symbol{\numexpr`a+\idx}}$} syndrome;
    } (msg[-2], syndrome[-2]);

    ["Recovery"]
    box {$\mathcal R$} (msg) | syndrome;
    discard syndrome;
  \end{yquant}
\end{tikzpicture}
         \end{codeexample}

      \subsection*{Circuit equations}
         \emph{The \texttt{groups} subpackage is new as of yquant~0.5.}

         
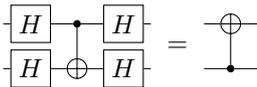
\begin{codeexample}
% \useyquantlanguage{groups}
\begin{tikzpicture}
   \begin{yquantgroup}
      \registers{
         qubit {} q[2];
      }
      \circuit{
         h -;
         cnot q[1] | q[0];
         h -;
      }
      \equals
      \circuit{
         cnot q[0] | q[1];
      }
   \end{yquantgroup}
\end{tikzpicture}
         \end{codeexample}

   \endwidetext
   \section{Quick guide}
      \subsection{General syntax}
         Every command in yquant may be prefixed by attributes (command\hyp specific and defined in the yquant manual~\cite{bib:yquantsrc}) or \TikZ{} options in square braces.
         After the attributes, the name of the command/gate has to be given; it is case\hyp insensitive.
         The name may then be followed by a value for this gate surrounded by curly braces.
         The value is followed by one or multiple target registers; optionally, and delimited by a pipe, one or multiple control registers may follow; again optionally and delimited by a tilde, one or more negative control registers may follow (i.e., conditioning on the bit not being set).
         Every command is ended by a semicolon.
         Whitespaces and line breaks (as well as ordinary \TeX{} comments) can be used at any time to make the code more readable.

      \subsection{Interaction with \TikZ}
         Every command can take the \texttt{name} attribute, which will assign a \TikZ{} node name to the shape.
         Additionally, you may at any time inject \TikZ{} or pure \TeX{} code.
         Consider the following example:

         \begin{codeexample*}
\begin{tikzpicture}
   \begin{yquant}
      qubit a[5];
      \foreach \x in {1, ..., 4} {
         \yquant
         [name=swapgate\x]
         swap a[(0, \x)];
      }
      \draw[red, ->] (swapgate1-0-s1) -- (swapgate2-0-s0) to[in=70] (swapgate3-0-s1.north east);
      cnot a[1, 4] | a[2, 3];
   \end{yquant}
\end{tikzpicture}
         \end{codeexample*}
         Here, we use the \tex!\foreach! command from \TikZ{} to produce multiple gates at once.
         Since using a \TeX{} command will stop the yquant engine, it has to be re\hyp started inside the loop with \tex!\yquant!.
         We then declare named gates in the loop.
         After the loop, we draw a line from the first (and only) multi\hyp qubit gate specified under the name \texttt{swapgate1}; and, more precisely, from its second element.
         The naming scheme which then leads to the full name \texttt{swapgate1-0-s1} is explained in detail in the manual.
         The line then goes to the next gate, but this time to the \emph{first} element in the first---and only---multi\hyp qubit part, and it finally ends at the lower part of the third gate, for which we even specify an anchor.

         After all these actions, another yquant gate follows.
         In principle, we would also have to say \tex!\yquant! before this last yquant command (and in fact, it does not harm to do so); but yquant takes care of this after all \TikZ{} \tex!\path!\hyp like operations.

      \subsection{Register declaration}
         Before using a register, a name has to be assigned by which it can be addressed at a later stage.
         yquant supports the declaration of vector registers: Apart from a name, the registers will get an additional zero\hyp based index, given in square braces.
         Upon declaration, the length of the register is given in the square braces.
         If those are omitted, a scalar register (length~$1$) is assumed by default.

         yquant supports registers of the types \texttt{qubit} (quantum, one line), \texttt{cbit} (classical, double\hyp line), \texttt{qubits} (quantum bundle, triple line---more common is to use \texttt{qubit} and a slash at the beginning), and \texttt{nobit} (no line, for discarded wires).

         For simple circuits, one might want to just use qubit registers without initial texts.
         In this case, use the \texttt{yquant*} environment (with a star); yquant will then create a qubit register whenever you refer to an unknown register name for the first time:
         \begin{codeexample}
\begin{tikzpicture}
   \begin{yquant*}
      x a;
      y a;
      cnot a | b;
   \end{yquant*}
\end{tikzpicture}
         \end{codeexample}
         yquant even supports discontiguous vector registers; refer to the full documentation for more information on this.

      \subsection{Register usage}
         In this subsection, we will assume that we declared a vector register with name \texttt{a} and length $7$, and a scalar register with name \texttt{b}, in this order.
         For more complicated cases (where, e.g., \texttt{a} might be continued after \texttt{b}), refer to the full documentation.

         A register is addressed by using its full name, i.e., the assigned name followed by its vector index in square braces: \texttt{a[3]} refers to the fourth register with name \texttt{a}.

         More than one register can be used by joining multiple full names in a comma\hyp separated list: \texttt{a[3], a[6]} will perform the given operation twice at the same time---once on the fourth, and once on the seventh register of \texttt{a}.
         Multiple registers might also be given as a range, which consists of a full name of the first, a dash, and the full name of the second register: \texttt{a[3]-a[6]} is equivalent to \texttt{a[3], a[4], a[5], a[6]}.
         The first name may be omitted, which then implicitly refers to the first register in the circuit: \texttt{-a[2]} is equivalent to \texttt{a[0], a[1], a[2]}.
         The last name may be omitted, which then implicitly refers to the last register in the circuit (at the time of using the register, as registers may be declared at any time): \texttt{a[5]-} is equivalent to \texttt{a[5], a[6], b}.

         Apart from using comma\hyp separated lists and ranges with full names, they may also be used inside the indices: \texttt{a[1, 3-6]} is equivalent to \texttt{a[1], a[3], a[4], a[5], a[6]}.
         If, within a range in an index the first number is omitted, this implicitly refers to the index zero; if the last index is omitted, this implicitly refers to the last index within this register (at the time of using the register): \texttt{a[-2]} is equivalent to \texttt{a[0, 1, 2]}; and \texttt{a[5-]} is equivalent to \texttt{a[5, 6]} (note in particular the difference to \texttt{a[5]-}, which is highlighted in yellow in the example circuit below).

         If a register is addressed without any index, this is equivalent to an open range: \texttt{a} means the same as \texttt{a[-]}, which in turn here is \texttt{a[0-6]}.
         Scalar registers are treated in the exact same way as vector registers of length~$1$; so instead of \texttt{b}, one might also say \texttt{b[0]} or \texttt{b[-]}.

         Finally, if instead of acting on multiple registers individually, a gate should act on the specified registers \emph{jointly} (i.e., a multi\hyp qubit gate), the relevant gates must be surrounded by parentheses: \texttt{(a[0], a[1])}, \texttt{(a[0]-a[1])}, \texttt{a[(0, 1)]}, and \texttt{a[(0-1)]} will all give a multi\hyp qubit gate on the first and second register with the name \texttt{a}.
         \texttt{(a)} is equivalent to \texttt{a[(-)]}, or \texttt{a[(0-6)]}; \texttt{(-)} is equivalent to \texttt{(a-b)} or \texttt{(a[0]-b)}.
         Multi\hyp qubit registers and single\hyp qubit registers can be mixed arbitrarily, as long as the gate in use supports multi\hyp qubit registers at all.

         All the addressing techniques described before hold in the same way for register targets, controls, and negative controls; but multi\hyp qubit registers can only be used as targets.

         The following circuit depicts the usages introduced above.

         \widetext
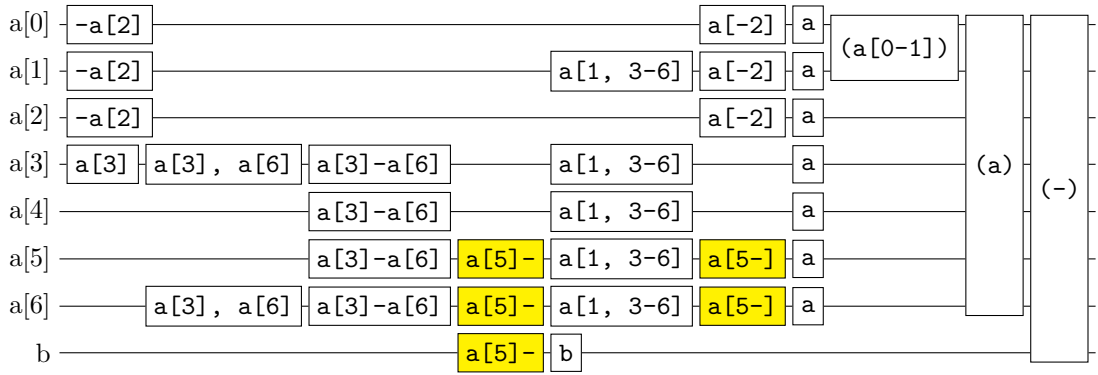
            \begin{codeexample**}
\begin{tikzpicture}
   \begin{yquant}[operators/every box/.append style={font=\ttfamily}, operator/minimum width=0pt]
      qubit a[7];
      qubit b;
      box {a[3]} a[3];
      box {a[3], a[6]} a[3], a[6];
      box {a[3]-a[6]} a[3]-a[6];
      box {-a[2]} -a[2];
      [fill=yellow]
      box {a[5]-} a[5]-;
      box {a[1, 3-6]} a[1, 3-6];
      box {a[-2]} a[-2];
      [fill=yellow]
      box {a[5-]} a[5-];
      box {a} a;
      box {b} b;
      box {(a[0-1])} (a[0-1]);
      box {(a)} (a);
      box {(-)} (-);
   \end{yquant}
\end{tikzpicture}
            \end{codeexample**}

   \section{List of gates}
      In this section, all gates natively shipped with yquant are listed.
      A detailed documentation of all gates and ways to define custom gates can be found in the full documentation.
      Note that all default appearances may be configured almost arbitrarily.

      \subsection{Visible gates}
         The gates listed here produce a distinct shape.
         \begin{itemize}
            \item \texttt{barrier} (vertical line)
            \item \texttt{box} (arbitrary text in a box)
            \item \texttt{correlate} (\tikz[baseline=-5.3mm] {\begin{yquant}[register/minimum height=2mm] qubit {} a[2]; correlate (a); \end{yquant}})
            \item \texttt{cnot} ($\oplus$)
            \item \texttt{dmeter} (\tikz[baseline=-3mm] {\begin{yquant} qubit {} a; dmeter a; \end{yquant}}), turns wire classical
            \item \texttt{h} (\fbox{$H$})
            \item \texttt{inspect} (arbitrary text, unboxed)
            \item \texttt{measure} (\tikz[baseline=-3.5mm] {\begin{yquant} qubit {} a; measure a; \end{yquant}}), turns wire classical
            \item \texttt{not} ($\oplus$)
            \item \texttt{phase} ($\bullet$)
            \item \texttt{slash} (\tikz {\begin{yquant} nobit a; slash a; \end{yquant}})
            \item \texttt{subcircuit} (nested circuit in a box)
            \item \texttt{swap} (\tikz[baseline=-5.3mm] {\begin{yquant}[register/minimum height=2mm] qubit {} a[2]; swap (a); \end{yquant}})
            \item \texttt{x} (\fbox{$X$})
            \item \texttt{xx} (\tikz[baseline=-5.3mm] {\begin{yquant}[register/minimum height=2mm] qubit {} a[2]; xx (a); \end{yquant}})
            \item \texttt{y} (\fbox{$Y$})
            \item \texttt{z} (\fbox{$Z$})
            \item \texttt{zz} (\tikz[baseline=-5.3mm] {\begin{yquant}[register/minimum height=2mm] qubit {} a[2]; zz (a); \end{yquant}})
         \end{itemize}

      \subsection{Pseudo-gates}
         The gates listed here have an impact on the circuit, but do not produce shapes themselves.
         \begin{itemize}
            \item \texttt{addstyle} (add \TikZ{} options to change the line style of a register)
            \item \texttt{align} (horizontally align the next gates in the targeted registers)
            \item \texttt{discard} (change a register into \texttt{nobit})
            \item \texttt{hspace} (manually insert whitespace of a given dimension in a register)
            \item \texttt{setstyle} (modify \TikZ{} options to change the line style of a register)
            \item \texttt{settype} (manually change the type of a register)
         \end{itemize}

      \subsection{Initializers and finalizers}
         \begin{itemize}
            \item \texttt{qubit}, \texttt{cbit}, \texttt{qubits}, \texttt{nobit} (create a new register of a given type)
            \item \texttt{init} (re-initialize a given register---also multi\hyp qubit registers, which will be connected by a curly brace---with a certain value)
            \item \texttt{output} (put an output value at the end of the given register---also multi\hyp qubit registers, which will be connected by a curly brace)
         \end{itemize}

   \section{Foreign language support: qasm}
      If you already have quantum circuits written in qasm, you can switch to yquant very quickly.
      This will lead to a better quality---lines now truly connect---and the possibility to change your circuits without running an external script.
      You may choose to inline your circuits directly into the \TeX{} file, but you may also leave them in external files.
      Refer to the full documentation to see how to use the macros \tex!\yquantimport! and \tex!\qasmimport!.

      If you want to use the qasm module, you must additionally say \tex!\useyquantlanguage{qasm}! after loading yquant in the preamble.
      Having done this, the teleportation circuit might for example also be written as
         \begin{codeexample*}
\begin{tikzpicture}
   \begin{qasm}
        qubit   q0
        qubit   q1
        qubit   q2

        h       q1     # create EPR pair
        cnot    q1,q2
        cnot    q0,q1  # Bell basis measurement
        h       q0
        nop     q1
        measure q0
        measure q1
        c-x     q1,q2  # correction step
        c-z     q0,q2
   \end{qasm}
\end{tikzpicture}
         \end{codeexample*}
         Note that here, compared to the same circuit written in yquant's own language, the two measurement gates are independent and thus not exactly aligned horizontally.
         This is due to the fact that circuits are not rendered using a tabular layout, but each wire individually keeps track of its own horizontal position.
         Consequently, the \texttt{nop} gate is just an approximation.

   \section{Publishing with yquant}
      At the moment, the arXiv uses a slightly updated TeXlive~2020, which means that yquant~0.3.2 is supported out\hyp of\hyp the\hyp box.
      Since there has been a lot of development, additional features and important bugfixes since then, it is recommended to upload the current version of yquant together with your publication.
      For every version since~0.4, the Releases section on GitHub~\cite{bib:yquantsrc} provides a single\hyp file version of this package (plus all subpackages and the corresponding documentation).
      The changelog in the full documentation reveals which features are missing compared to the most current version.

      Since even \TeX\hyp based journals usually use very outdated versions of \TeX, the circuits should be externalized (e.g., using the standalone~\cite{bib:standalone} package or \TikZ's externalization) into separate PDF files before submission.

      Also note that old \TeX{} versions will maybe raise some compilation errors due to macro protection.
      For example, it is not necessary to \tex!\protect! matrix environments (including the ampersands) in \texttt{box} gates in modern distributions, but older ones will require this.

      Always make sure to keep an up\hyp to\hyp date \TeX{} distribution locally.
      There is currently a lot of development happening both in the kernels as well as in \TikZ, and new yquant versions may not work with old kernels or packages.

   \RaggedRight
   \bibliography{ms}
\end{document}